\documentclass[a4paper,11pt]{article}

\title{\textbf{Infrared Thermography of Complex 3D Printed Components}}
\author{\textbf{BY:}\\ \\Wong Ming Hin\\ \\NUS High School\\ \\ \\\textbf{External Mentor}\\ \\Henry Goh Kok Hin\\ \\A*STAR}
\date{3rd Jan 2018}

\usepackage{amsmath} 
\usepackage{graphicx} 
\usepackage{float} 
\usepackage{subcaption} 
\usepackage[margin=2.5cm]{geometry}
\usepackage{ulem} 
\usepackage[colorlinks=true,linkcolor=black, citecolor=blue]{hyperref} 

\graphicspath{ {Images/} }
\newcommand{\Henry}[2]{\textcolor{red}{#1\sout{#2}}}

\begin{document}

%
%
\maketitle

\begin{abstract}
The possibility of using Infrared Lock-In Thermography (LIT) to estimate the thickness of a sample was assessed and shown to be accurate up to 1.8mm. LIT is a technique involving heating samples with halogen lamps with varying intensity over time. The intensity is defined by sinusoidal functions. LIT was conducted on samples of varying thickness, gradient, and shape. The Lock-In phase signals were calculated, and a database was then created with the data obtained and was used to estimate the thickness based on the original phase signal. A relationship between gradient and phase signal was also shown based on our data, contrary to current findings in existing literature.
\end{abstract}

\break
\section{Introduction}

Infrared Thermography is a non-destructive testing technique which detects the infrared range of the electromagnetic spectrum to produce thermograms. It is split into active and passive thermography where active thermography involves a sample directly heated by a heat source. Lock-In Thermography (LIT) is one example of active thermography which involves heating samples with varying intensity over time. This intensity is defined by sinusoidal functions with constant frequencies (Lock-In frequency). Compared to steady state thermography, this has a greater signal to noise ratio \cite{Straube2011SEMSC}. The phase image output by this technique can highlight defects and other anomalies at different depths in a sample using different lock in frequencies \cite{Delanthabettu2015QITJ, Huth2002BOOK, Wallbrink2007JAP}. It is also relatively insensitive to non-uniform heating \cite{Spiessberger2010NDTA}. While this phase signal is very useful in highlighting defects and anomalies, it is not in direct correlation with physical values such as thickness, emissivity, and other material properties \cite{Spiessberger2010NDTA}. 
\\
\\
Hence, the aim of this project is to assess the possibility of mapping the thickness of a sample to the signal obtained from LIT. This has been done multiple times with coating thickness \cite{Wu1996BOOK,Sakagami2002IPT,Ranjit2017JKS} as well as material thickness itself \cite{Spiessberger2010NDTA}. However, tested samples are either in large steps, or a simple wedge with a constant gradient. In this project, we have investigated both curved samples as well as linear samples to test the effect that the gradient has on the signal as well. This will allow the technique to identify anomalies in samples where the thickness values are known. For sample regions where the signals deviate from expected range of values, it is possible that defects exist. Hence, this technique will enhance the identification of defects in non-destructive testing.

\section{Lock-In Thermography}

Lock-In Thermography has many applications in testing for subsurface defects as it is non-destructive and has a high signal to noise ratio allowing it to amplify the defect signals, making it more visible in the phase image. This is done using a 2 – phase Lock-In amplifier, which uses the in-phase and out-of-phase signal, to calculate the amplitude, as well as the phase image \cite{Huth2002BOOK, Breitenstein2004BOOK}. The in-phase signal can be found by Equation~\ref{Eq:S0}, while the out-of-phase signal can be found by Equation~\ref{Eq:Sm90} 

\begin{equation}
S_0 (i,j) = \sum_{t=0}^{T}(sin(2\pi ft)*F(i,j,t))
\label{Eq:S0}
\end{equation} 

\begin{equation}
S_{-90} (i,j) = \sum_{t=0}^{T}(-cos(2\pi ft)*F(i,j,t))
\label{Eq:Sm90}
\end{equation}

where $f$ is the Lock-In frequency, $F(t)$ is the intensity reading of a camera pixel $(i,j)$ at a certain point in time $t$, and $T$ is the total time taken per measurement. The amplitude and phase can then be calculated with Equations 3 and 4 respectively \cite{Huth2002BOOK, Breitenstein2004BOOK}

\begin{equation}
A= \sqrt {S_0^2+S_{-90}^2}
\end{equation}

\begin{equation}
\phi = \arctan(\frac {-S_{-90}}{S_0})
\end{equation} 

\section{Experiments}

\begin{figure}[H]
 
\begin{subfigure}{0.5\textwidth}
\includegraphics[width=\linewidth,keepaspectratio]{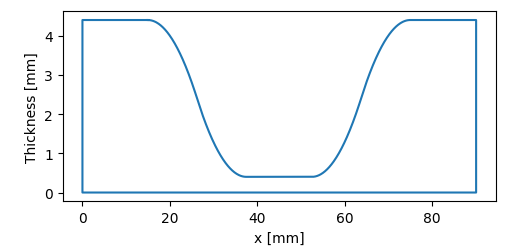} 
\caption{Cross section of a curved sample, sample 1}
\label{fig:sample}
\end{subfigure}
\begin{subfigure}{0.5\textwidth}
\includegraphics[width=\linewidth,keepaspectratio]{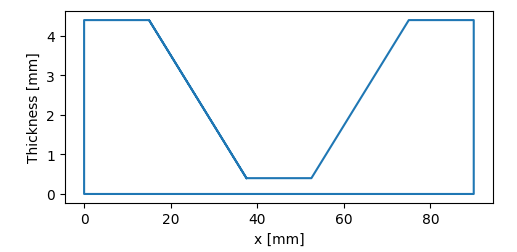}
\caption{Cross section of a linear sample, sample 6}
\label{fig:linearsample}
\end{subfigure}
\caption{Design of Samples}
\end{figure}

We designed 5 different samples similar to the one in Figure~\ref{fig:sample} which varies in thickness from 0.4mm to 4.4mm. Different samples varied in length and hence gradient. While gradient has not been known to affect the phase signal, it is later shown that it does influence the signal. Hence, 8 more linear samples with constant gradient were designed similar to the one Figure~\ref{fig:linearsample}.

Buffer areas 15mm long were designed in the sample to reduce the effect that faster cooling on the edges would have on our results. The samples were then 3D printed with ABS plastic with layer height of 0.2mm. Details of our samples are found in Table \ref{tbl:samples}.

\begin{table}[h]
\begin{center}
\caption{Sample Details}
\label{tbl:samples} 
\begin{tabular}{|c|c|c|c|c|} \hline

Sample & Length [mm] & Max gradient & Width [mm] & Gradient Type \\ \hline
1 & 90 & 0.356&56&Quadratic \\
2 & 105 & 0.267&56&Quadratic \\
3 & 75 &0.533 &56&Quadratic \\
4 & 80 & 0.457 &56&Quadratic \\
5 & 85 & 0.400&56&Quadratic \\ \hline
6 & 90 & 0.178 &56&Linear \\
7 & 105 & 0.133&56& Linear \\
8 & 75 & 0.267&56& Linear \\
9 & 80 & 0.229&56&Linear \\
10 & 85 & 0.200&56 & Linear \\ 
11 & 60 &0.533&56& Linear \\
12 & 65 &0.400&56 & Linear \\
13 & 70 &0.320&56 & Linear \\ \hline

\end{tabular}
\end{center}
\end{table}

\begin{figure}[H]
\centering
\includegraphics[height=7cm]{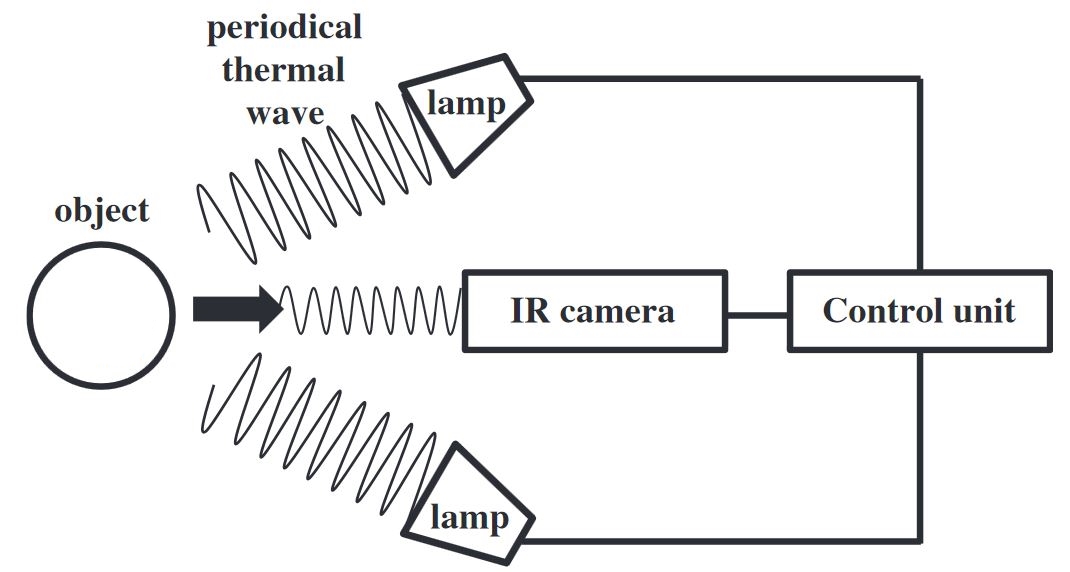}
\caption{Setup for Lock-In Thermography}
\label{fig:setup}
\cite{Kim2014IPT}
\end{figure}

Our experiment was setup according to Figure \ref{fig:setup}. 4 halogen lamps were used to minimise the effects of uneven heating. The halogen lamps were controlled by a custom-built unit that heated the samples with a Lock-In frequency of 0.01 Hz. The sample was heated over a period of 200s with an IR camera (Xenics Gobi640) capturing thermograms at 1 frame per second. Each image had a resolution of 320 by 240 pixels. Resultant phase images from Lock-In calculations were mapped to corresponding coordinates of samples, using a custom written algorithm.

\section{Results and Discussion}
\begin{figure}[H]
\begin{subfigure}{0.5\textwidth}
\includegraphics[width=\linewidth,height=45mm]{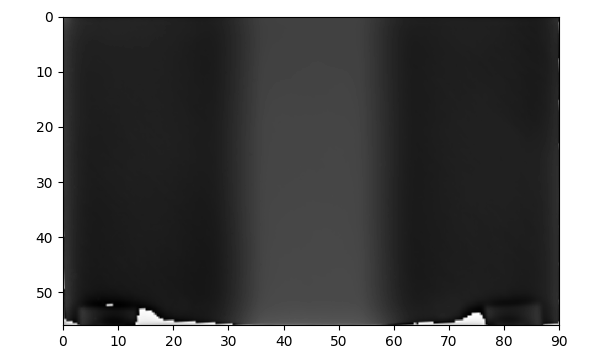}
\caption{Phase Image of sample 1 }
\label{fig:phase}
\end{subfigure}
\begin{subfigure}{0.5\textwidth}
\includegraphics[width=\linewidth,height=45mm]{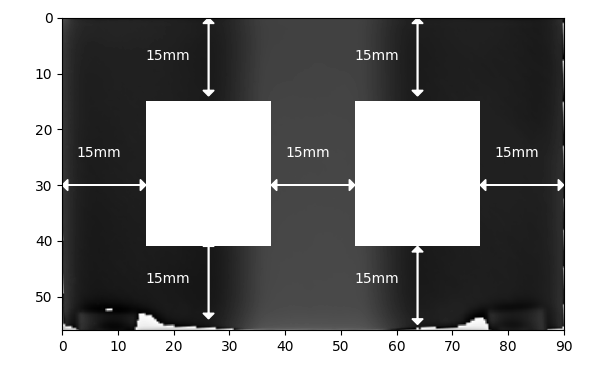}
\caption{Area of Interest of sample 1}
\label{fig:aoi}
\end{subfigure}
\caption{}
\end{figure}

Figure \ref{fig:phase} is the phase image of sample 1 cropped from a larger image obtained from the Lock-In algorithm. This shows that a difference in thickness can be seen from the phase image. Figure \ref{fig:aoi} shows the area of interest which we analysed further as this would reduce the effect that faster cooling on the edges would have on our results by isolating the centre of the sample out. The dimensions shown are in millimetres

\begin{figure}[H]
\centering
\includegraphics[height=7cm,keepaspectratio]{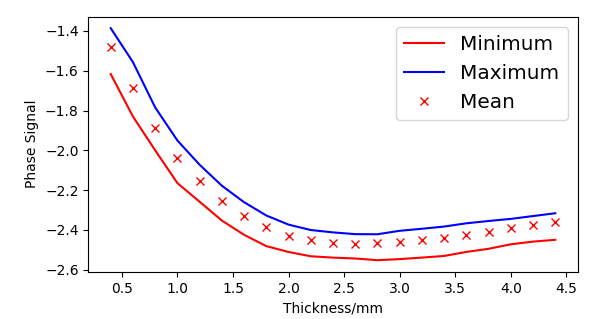}
\caption{Graph of Phase Signal against Thickness for sample 1}
\label{fig:pt}
\end{figure}

This allowed us to plot the phase signal of the pixel, against the thickness of that pixel in Figure \ref{fig:pt}, by collecting data in bulk and generating statistics. It clearly shows the trend of the phase against the thickness. However, there is a rather large deviation of the maximum and minimum possible phase signal for each thickness. This is likely due to a non-uniform heating which affects the phase signal.

\begin{figure}[H]
\centering
\includegraphics[width=121mm, height=80mm]{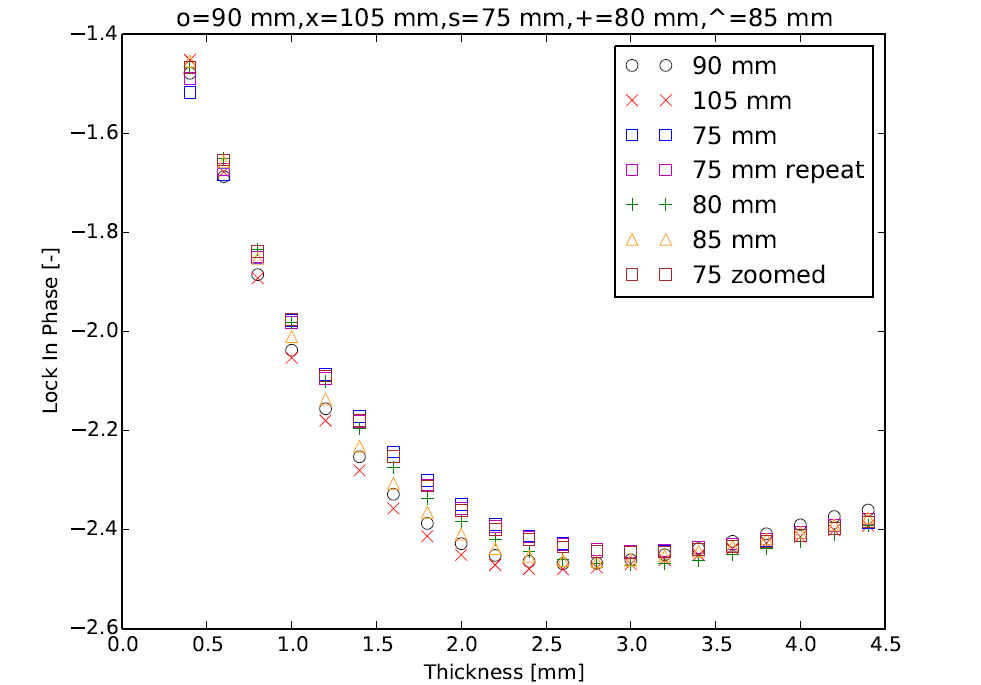}
\caption{Mean phase signal against Thickness for all curved samples}
\label{fig:csamp}
\end{figure}

However, when all 5 curved samples were plotted as seen in Figure \ref{fig:csamp}, a trend between the phase signal and the gradient of the sample can be seen. Figure \ref{fig:csamp} shows that and increasing length of the sample (hence decreasing the gradient), cause a decrease in in the phase signal. This trend can be more clearly seen at thicknesses of around 1.5mm to 2.0mm. Figure \ref{fig:csamp} also shows that the camera resolution had negligible effects on the phase signal (75 zoomed), and the repeatability of the setup is high (75 mm repeat).
 
\begin{figure}[H]
\centering
\includegraphics[width=121mm, height=80mm]{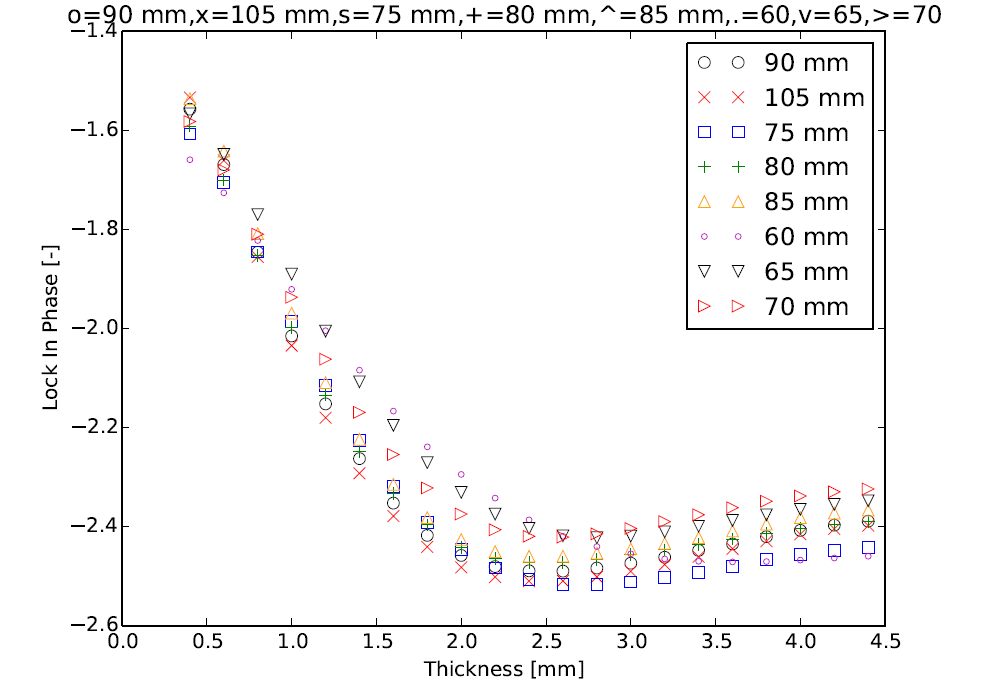}
\caption{Mean phase signal against Thickness for all linear samples }
\label{fig:lsamp}
\end{figure}

In order to confirm that the thickness and gradient were the dominant variables that affected the phase signal, we followed up with 8 linear samples with constant gradients and plotted results those samples. As can be seen in Figure \ref{fig:lsamp}, the similar trends were observed, further confirming that the gradient affects the phase signal. It was found that the phase profiles for a specific thickness and gradient match rather well between curved and linear samples. However, due to the camera resolution of 0.3mm per pixel, as well as steep changes in gradients in curved samples, the thickness mapped onto the pixel coordinate may have contained systematic errors resulting in a few poor matches between curved and linear samples. These phase profiles can be found in the appendix.

\begin{figure}[H]
\begin{subfigure}{\textwidth}
\center
\includegraphics[width=0.7\linewidth]{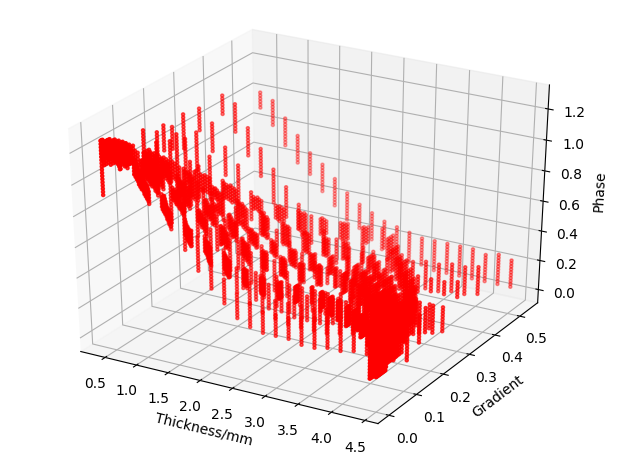}
\caption{3D database with all samples}
\end{subfigure}
\begin{subfigure}{0.5\textwidth}
\includegraphics[height=60mm,keepaspectratio]{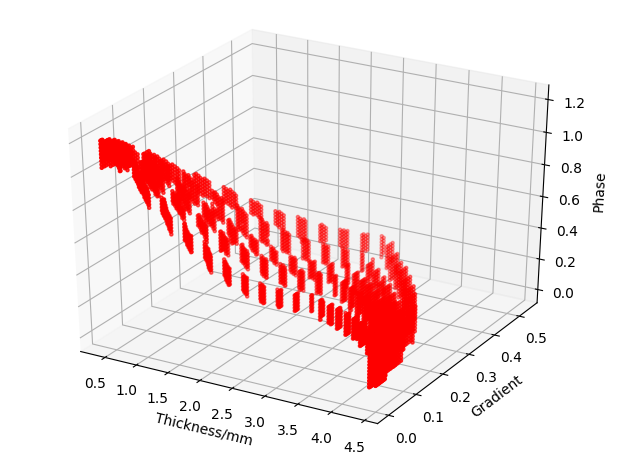}
\caption{3D database with curved samples}
\end{subfigure}
\begin{subfigure}{0.5\textwidth}
\includegraphics[height=60mm,keepaspectratio]{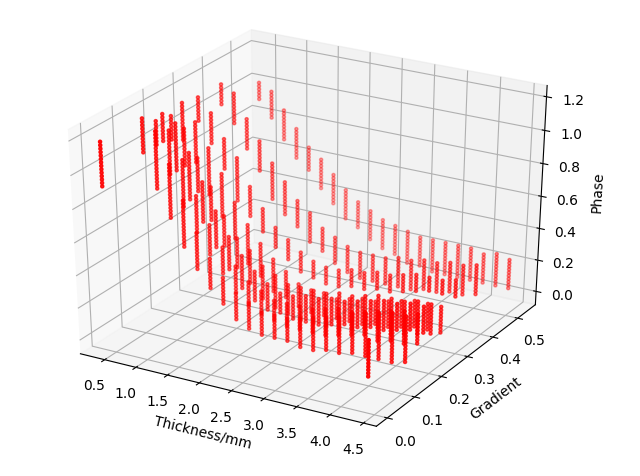}
\caption{3D database with linear samples}
\end{subfigure}
\caption{3D database, comparing different samples}
\label{fig:database}
\end{figure}

Assuming that the phase signal is a function of multiple dimensions, a customised framework was created in Python to enable the characterisation in multiple dimensions. However, the current study is limited in scope, and only includes thickness and gradient of ABS samples. Using this data, we created a 3D database with the probability of phase signals, given a particular thickness and gradient. The data points for this database are shown in  Figure \ref{fig:database}. 

\begin{figure}[H]
\begin{subfigure}{0.5\textwidth}
\includegraphics[width=\linewidth,height=45mm]{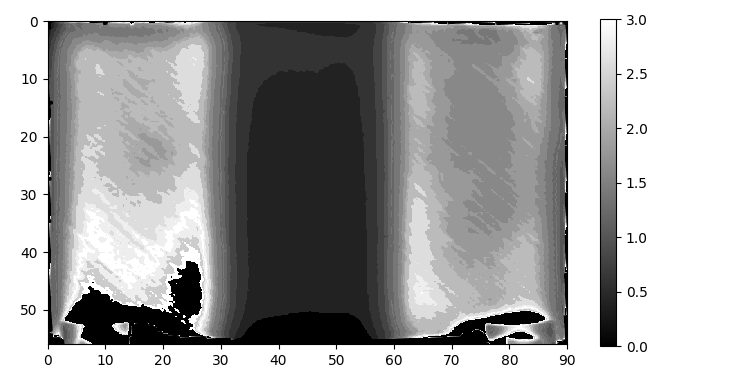}
\caption{Thickness Map \Henry{draw a line to indicate selected profile, label it}{}}
\label{fig:firstguess}
\end{subfigure}
\begin{subfigure}{0.5\textwidth}
\includegraphics[width=\linewidth,height=45mm]{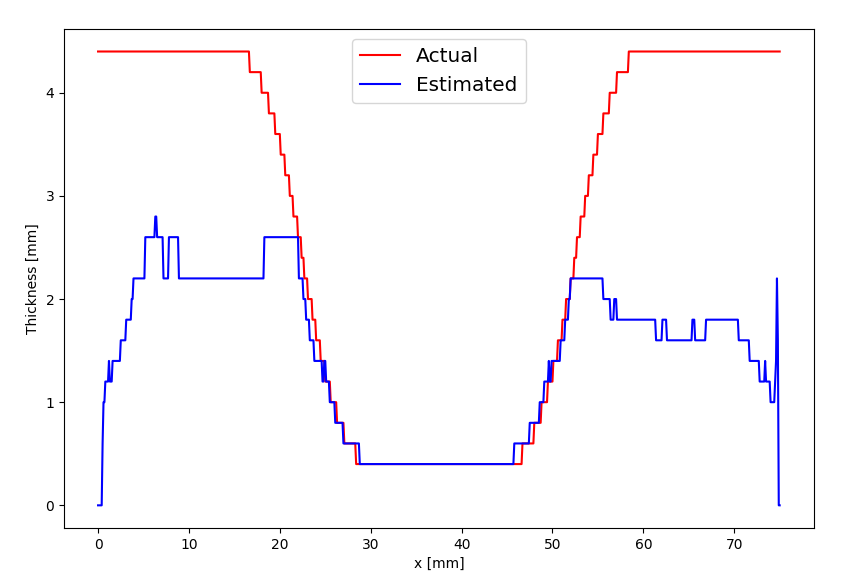}
\caption{Estimated against Actual Thickness}
\label{fig:est}
\end{subfigure}
\caption{}
\end{figure}

This database allowed us to map thickness back onto the sample using the phase signal and the database, by matching the thickness with the thickness of highest probability, given a phase value. An example of the thickness map for sample 3 can be seen in Figure \ref{fig:firstguess}. The corresponding actual and mapped thickness at the centre part of the sample is shown in Figure \ref{fig:est}

From Figure \ref{fig:firstguess}, the effects of uneven heating can be observed where the bottom half and top half of the sample were not heated equally. Similarly, the stands holding the sample also resulted in greater cooling of the lower portion of the sample. Also, the edges can be seen to cool much faster as well hence the dip in estimated thickness in Figure \ref{fig:est}

Our results show that we are able to map the thickness of a sample with a known material accurately to 0.2mm for thicknesses from 0.4mm to 1.8mm. Also, we have found that the gradient of a sample affects the phase signal which differs from the literature. This is a possible area which can be further researched to greater depth. However, our experiments are limited by the resolution of the camera (0.3mm), which may have caused inaccurate mapping of the thickness to the phase. Conducting more experiments at higher resolution on the same camera would reduce the systematic errors in mapping coordinates to thickness values. By conducting experiments on more samples with different gradients, as well as using different Lock-In frequencies, the database will show a more complete picture of the signal characteristics, which would allow us to pinpoint the thickness of any given sample more accurately. This will also facilitate the identification of defects, as they will cause deviations in the phase signals away from expected range of values at known thickness and gradient. This work could also be further developed using other materials as well, to understand the effects of material properties on the signals, which will enable material identification on known geometries. As it is expected that material properties and Lock-In frequencies will affect the phase signals, further studies would enable the creation of a multidimensional database (more than 5 dimensions). There is potential that such a database can be used for automated feature/material identification from measurements.

\section{Conclusion}

This project has shown the ability of LIT in mapping the thickness of a sample of ABS accurately up to 1.8mm. A 3D database has been generated from the data obtained for ABS with Lock-In frequency of 0.01Hz. This database has shown a relationship between the gradient and phase signal of a sample which has not been shown previously. Also, further studies can be conducted at different Lock-In frequencies on different materials, to create a multidimensional database. This will enable the identification of defects in samples where the thickness and materials are known. It could also possibly be used to identify materials given known sample geometry. Further studies in building the database will enhance and add capabilities in LIT, for possible automated feature and material identification.

\section*{Acknowledgements}
I would like to give special thanks to Dr Henry Goh Kok Hin from A*STAR for his time and guidance through this project

\break

\bibliographystyle{ieeetr}
\bibliography{reportbib}

\break
\section*{Appendix}

\begin{figure}[h]
\begin{subfigure}{0.5\textwidth}
\includegraphics[width=\linewidth, height=4.1cm]{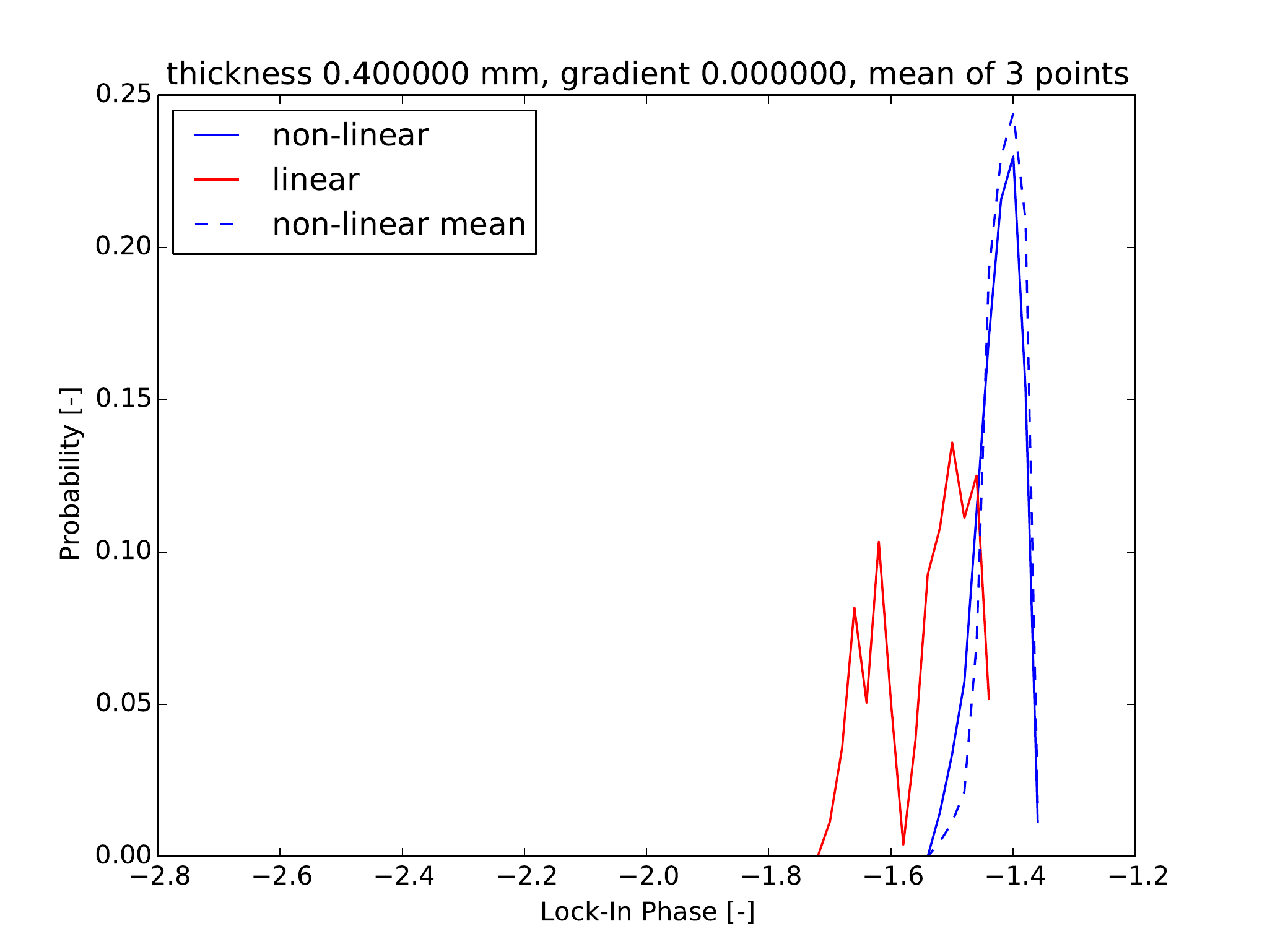}
\end{subfigure}
\begin{subfigure}{0.5\textwidth}
\includegraphics[width=\linewidth, height=4.1cm]{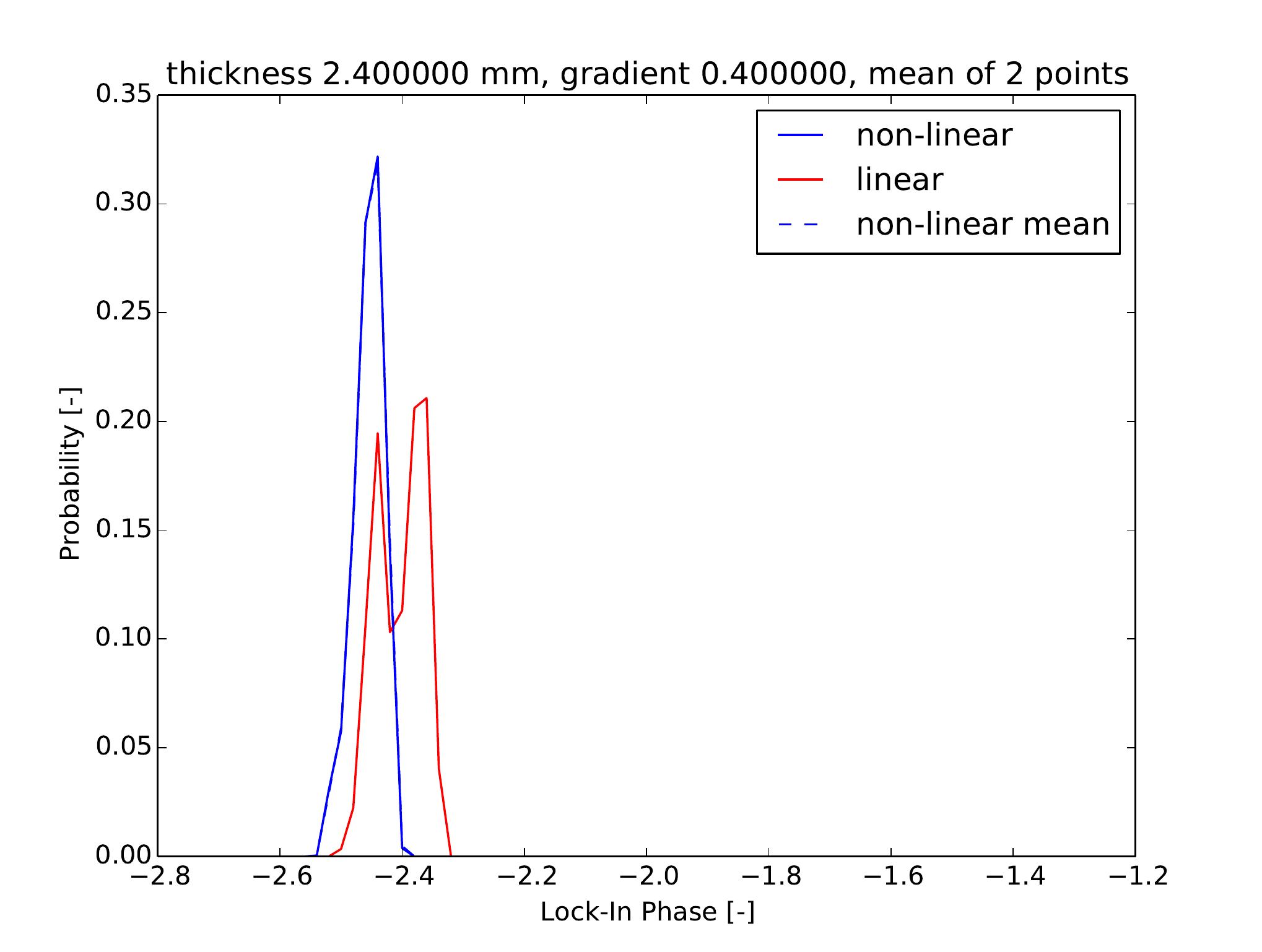}
\end{subfigure}
\caption{Poor match between Linear and Curved samples}
\end{figure}

\begin{figure}[h]
\begin{subfigure}{0.5\textwidth}
\includegraphics[width=\linewidth, height=4.1cm]{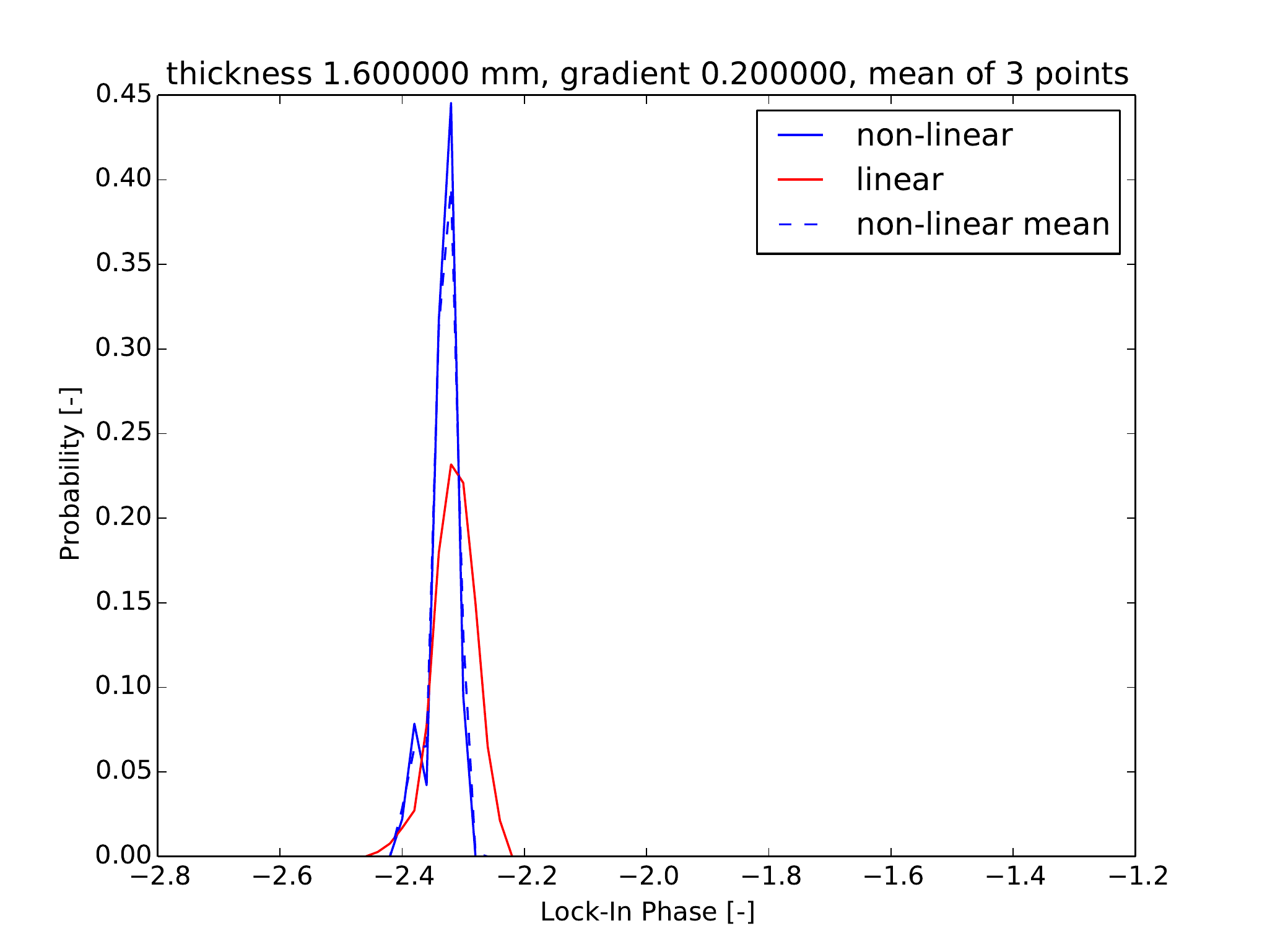}
\end{subfigure}
\begin{subfigure}{0.5\textwidth}
\includegraphics[width=\linewidth, height=4.1cm]{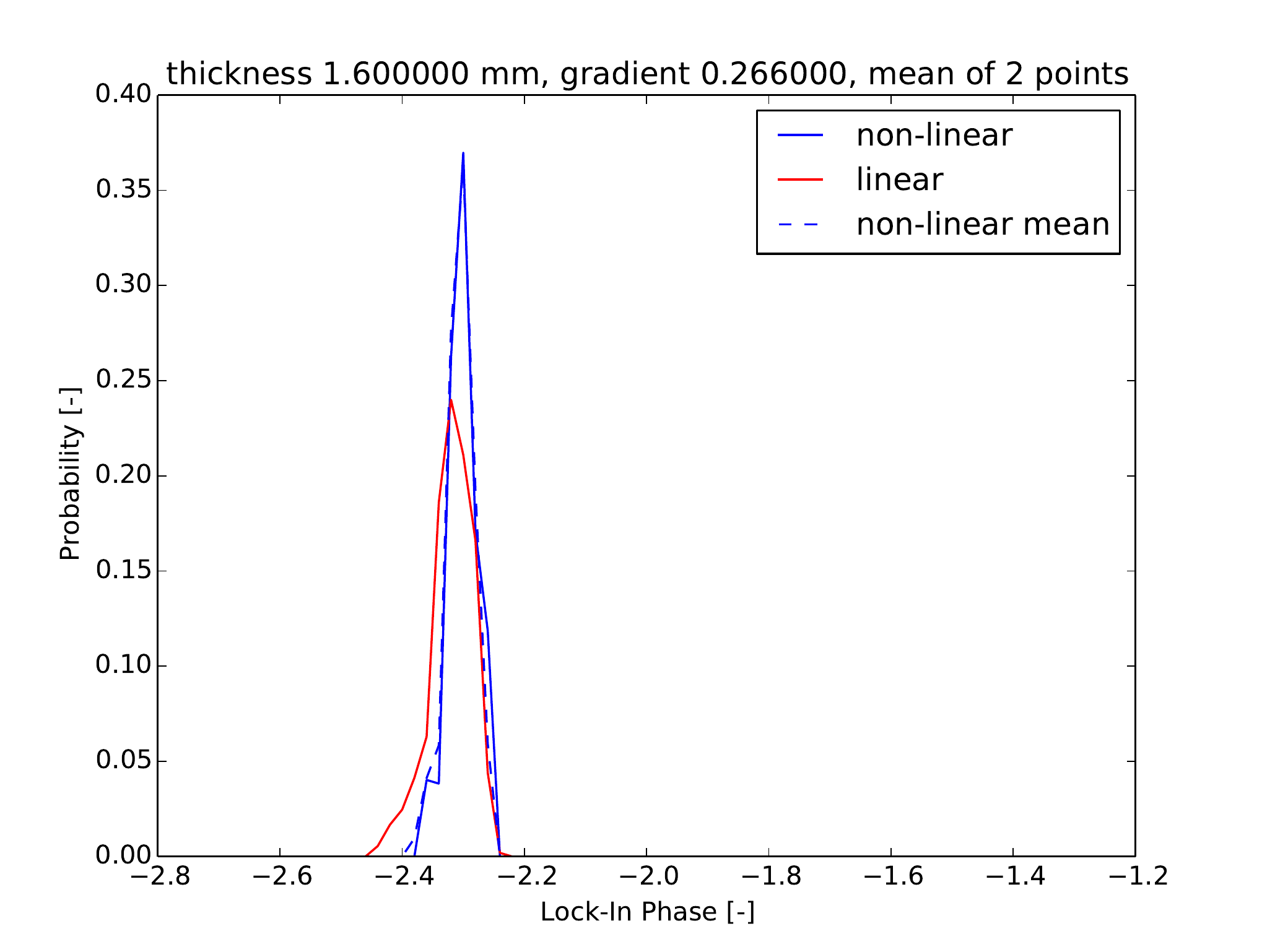}
\end{subfigure}
\begin{subfigure}{0.5\textwidth}
\includegraphics[width=\linewidth, height=4.1cm]{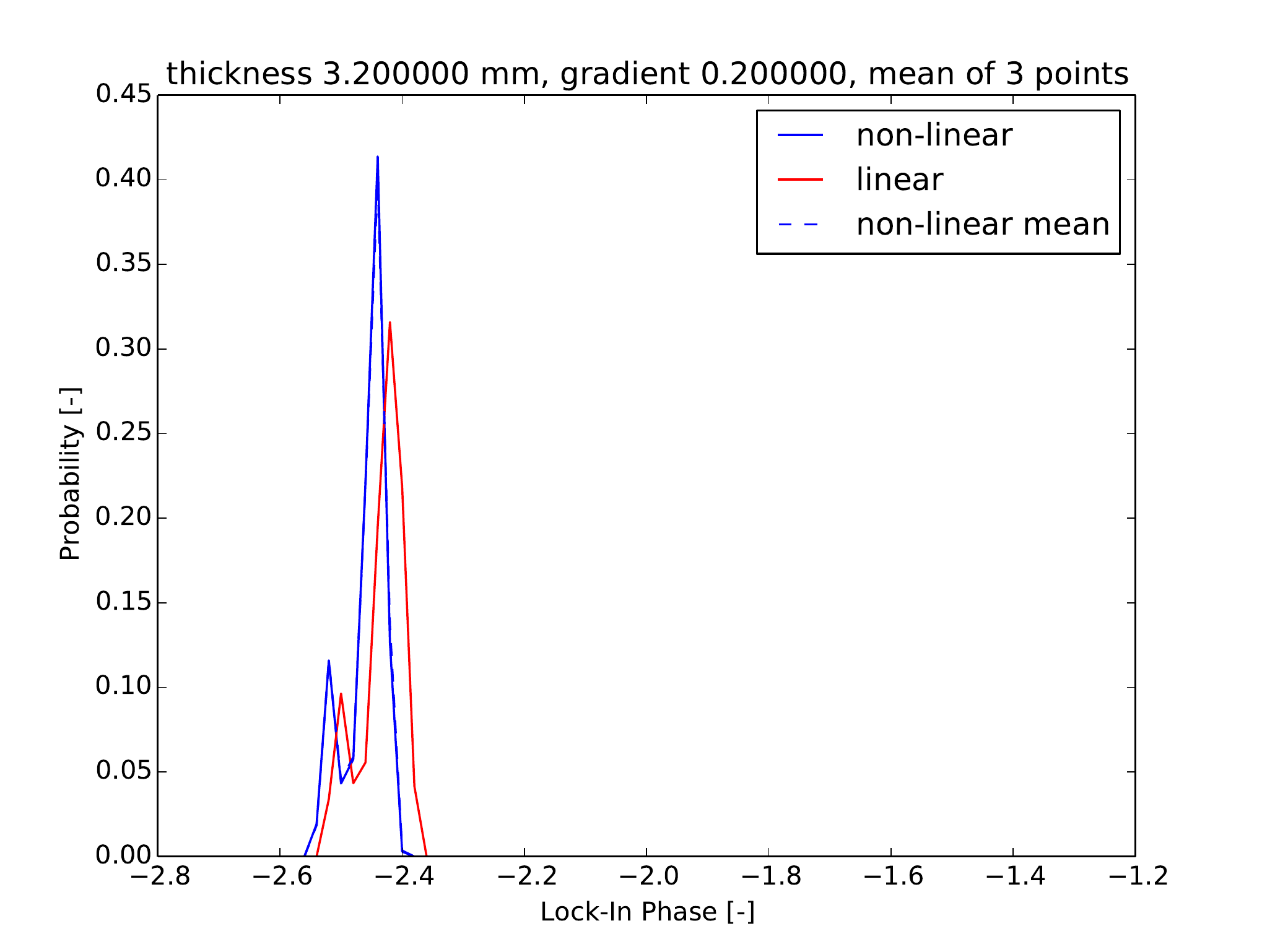}
\end{subfigure}
\begin{subfigure}{0.5\textwidth}
\includegraphics[width=\linewidth, height=4.1cm]{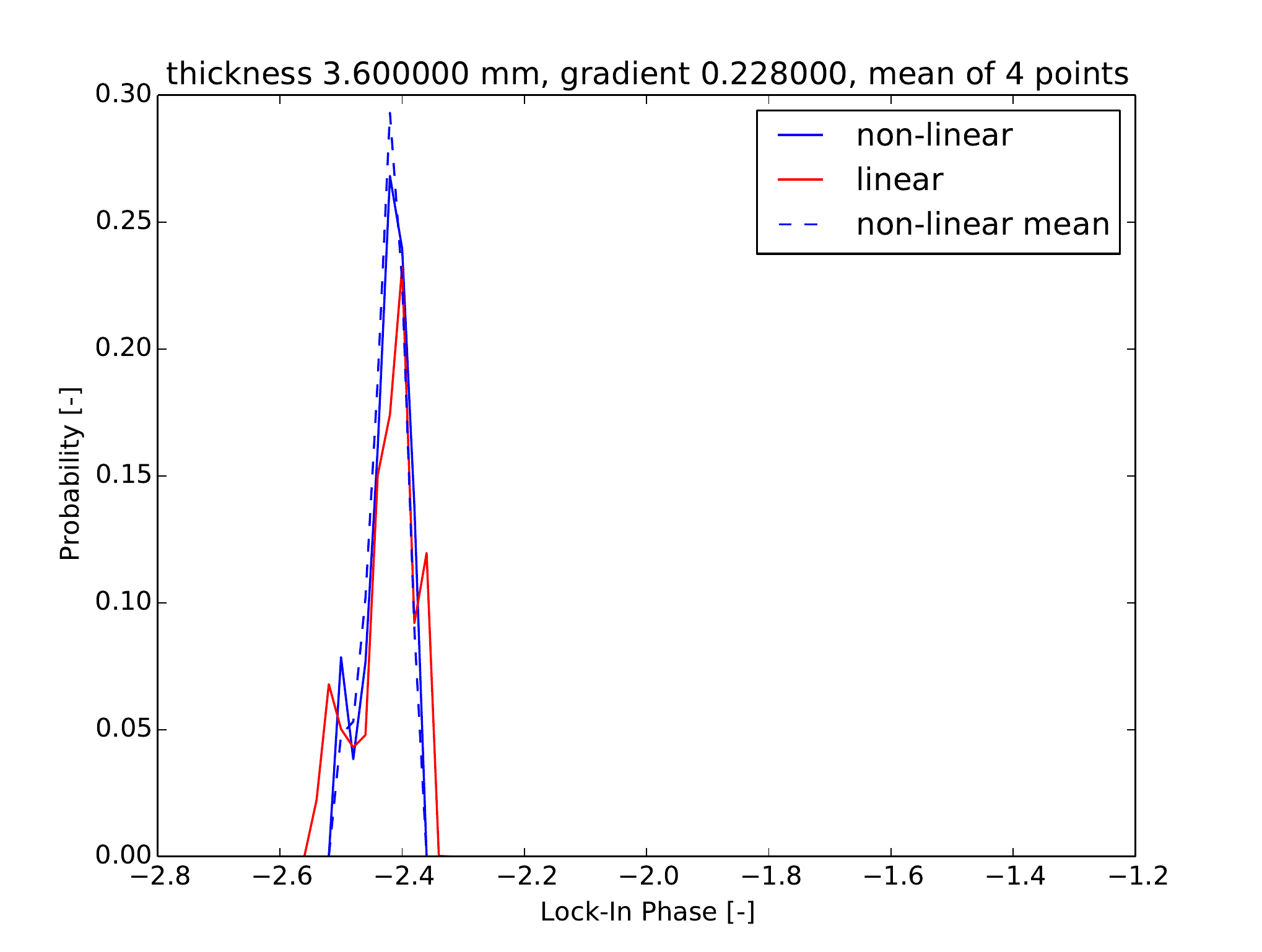}
\end{subfigure}
\begin{subfigure}{0.5\textwidth}
\includegraphics[width=\linewidth, height=4.1cm]{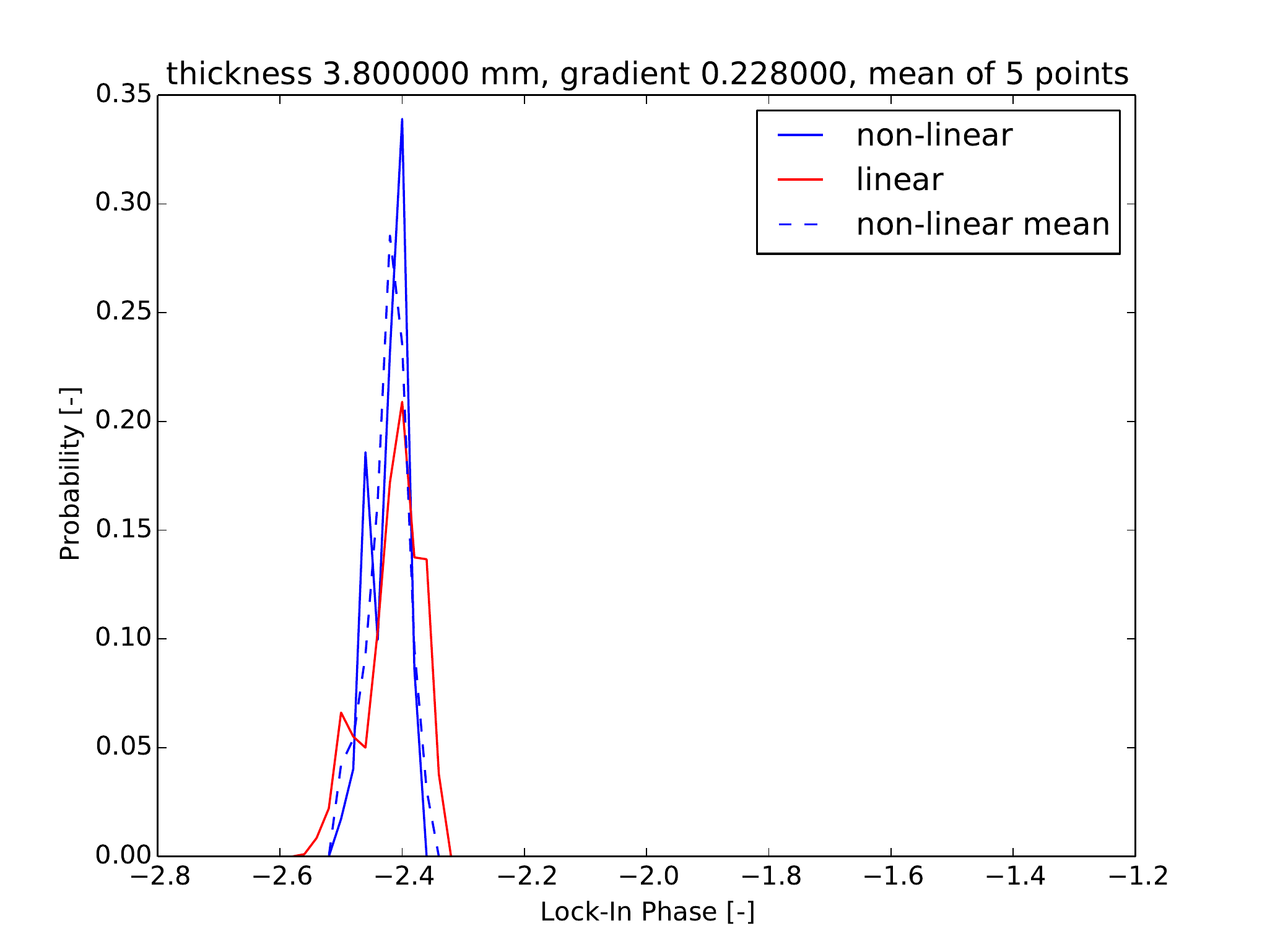}
\end{subfigure}
\begin{subfigure}{0.5\textwidth}
\includegraphics[width=\linewidth, height=4.1cm]{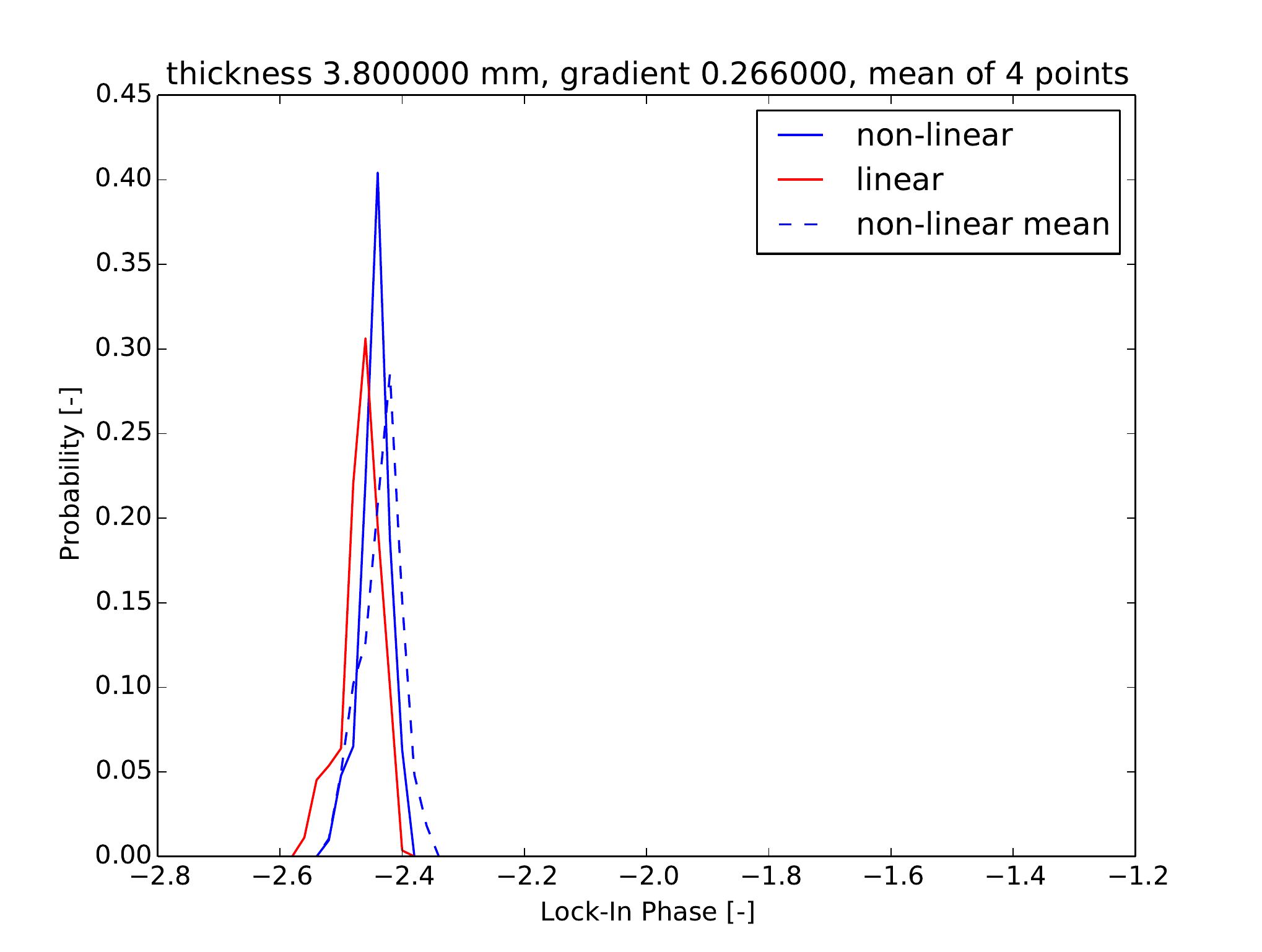}
\end{subfigure}
\caption{Good match between Linear and Curved sample}
\end{figure}


\end{document}